\documentclass[twocolumn,aps,showpacs,preprintnumbers,%
nofootinbib,superscriiptaddress]{revtex4}
\usepackage{graphicx}
\usepackage{dcolumn}
\usepackage{bm}
\begin{document}

\title{ENHANCED OPTICAL COOLING OF ION BEAMS FOR LHC\footnote{Supported by RFBR
under grant No 05-02-17162 and by NSF.\\$^\#$ bessonov@x4u.lebedev.ru}}

\author{E.G.Bessonov$^\#$,
M.V.Gorbunkov, Lebedev Phys. Inst. RAS, Moscow, Russia,
A.A.Mikhailichenko, Cornell University, Ithaca, NY, U.S.A.}

\begin{abstract}
The possibility of the enhanced optical cooling (EOC) of Lead ions in LHC is
investigated. Non-exponential feature of cooling and requirements to the
ring lattice, optical and laser systems are discussed. Comparison with
optical stochastic cooling (OSC) is represented.\end{abstract}

\date{\today}


\maketitle

\section{Introduction}

In original OSC with usage of quadrupole wiggler as a pickup [1], particles
with small betatron amplitude do not make an input into signal generation
(radiation), so they are not heating the beam. In contrast, in EOC method
such selective action achieved by usage of movable screens. These screens
located on image plane of optical system having radiating beam as source.
Motion realized with the help of fast electro-optical elements driven by
external voltage. As a result of this selection the ions with extreme
deviations of dynamic variables keep the neighboring ions undisturbed in the
first approximation. By this way the number of the particles in the
bandwidth, which defines the damping time can be reduced drastically. Some
detailed schemes of EOC were suggested in [2]--[4]. Below we consider EOC of
fully stripped Lead ions in LHC as example.

\section{THE SCHEME OF COOLING}

The EOC method uses a pickup undulator and one or more kicker undulators
installed in different straight sections of a storage ring. The distance
determined by a betatron phase advance $(2p - 1)\pi $ between the pickup and
the first kicker undulator and $2{p}'\pi $ between each of the following
kicker undulators; where$p,\;{p}'$ = 1, 2, 3... Undulator Radiation Wavelets
(URW), emitted by ions in the pickup undulator, transferred by optical
system to the movable screen located on the image plane. Here the
undesirable part of radiation, corresponding to small betatron amplitudes,
is cut. Residual fraction or URW amplified in optical amplifier and pass
together with the ions through the followed kicker undulators.

\section{THE RATE OF COOLING}

The change of the square of the amplitude of betatron oscillations of an
ion, caused by sudden energy change $\delta E$ in a kicker undulator is
determined in smooth approximation by

\begin{equation}
\label{eq1}
\delta A_x^2 = - 2x_{\beta ,k} \delta x_\eta + (\delta x_\eta )^2,
\end{equation}

\noindent
where $x_{\beta ,k} $ is the ion deviation from it's closed orbit in the
kicker undulator; $\delta x_\eta = \eta _x \beta ^{ - 2}(\delta E / E)$ is
the change of it's closed orbit position; $\eta _x $is the dispersion
function in the storage ring; $\beta $ is the normalized velocity. In the
approximation $\vert \delta x_\eta \vert < 2\vert x_{\beta ,k} \vert < 2A_x
$ both the betatron amplitude and the position of the closed orbit will be
decreased, if the values$x_{\beta ,k} < 0$, $\delta x_\eta < 0$. It follows
that to cool the ion beam the screen in the optical system must open the
pass for URWs emitted by extreme ions entering the pickup undulator with
higher energy and betatron deviations $x_{\beta, p} > 0$from theirs
closed orbits. After that the screen will open images of ions with lower and
lower energies until the optical system must be switched off. Then the
cooling process can be repeated. So the EOC is going simultaneously both in
the longitudinal and transverse degrees of freedom.

Optical lengths between pickup and kicker undulators should be picked up so
that to inject ions in the kicker undulators at decelerating phases of their
own URWs.

The total energy of the undulator radiation (UR) emitted by a relativistic
ion traversing an undulator with magnetic field $B$ is given by

\begin{equation}
\label{eq2}
E_{tot} = \textstyle{2 \over 3}r_i^2 \overline {B^2} \gamma ^2L_u ,
\end{equation}

\noindent
where $\overline {B^2} $ is an average square of magnetic field along the
undulator period $\lambda _u $; $r_i = Z^2e^2 / M_i c^2$ is the classical
radius of the ion; $e$, $M_i $ are the electron charge and ion mass
respectively; $Z$is the atomic number, $L_u = M\lambda _u $; $M$is the
number of undulator periods; $\gamma $ is the relativistic factor. For a
plane harmonic undulator $\overline {B^2} = B_0^2 / 2$, where $B_{0}$ is the
peak of the undulator field. For helical undulator $\overline {B^2} = B_0^2
$. The spectrum of the first harmonic of the UR is $dE_1 / d\xi = E_1 f(\xi
)$, where $E_1 =  \quad E_{tot} /  \quad (1 + K^2)$, $K = Ze\sqrt {\overline {B^2}
} \lambda _u / 2\pi M_i c^2$, $f(\xi ) = 3\xi (1 - 2\xi + 2\xi ^2),$
$\xi =  \quad \lambda _{1,\min } / \lambda _1 $, $\lambda _{1\min } = \lambda _1
\vert _{\theta = 0} $, ($0 \le \xi \le 1)$, $\int {f(} \xi )d\xi = 1$, $M >
> 1$, $\lambda _1 = \lambda _u (1 + K^2 + \vartheta ^2) / 2\gamma ^2$is the
wavelength of the first harmonic of the UR, $\vartheta = \gamma \theta $;
$\theta $, the azimuth angle.

The number of the equivalent photons in the URW in the suitable for cooling
frequency range $(\Delta \omega / \omega )_c = 1 / 2M$ and angular range
$\Delta \vartheta = \sqrt {(1 + K^2) / 2M} $

\begin{equation}
\label{eq3}
N_{ph} = \Delta E_1 / \hbar \omega _{1\max } = \pi \alpha Z^2K^2,
\end{equation}

\noindent
where $\Delta E_1 = (dE_1 / d\omega )\Delta \omega = 3E_{tot} / 2M(1 +
K^2)$, $\omega_{1\max } = 2\pi c / \lambda _{1\min } $, $M = L_u (1 +
K^2) / 2\gamma ^2\lambda _{1\min } $. An aperture or filters must be used in
the optical system to select a portion of URW in this frequency range for
resonance interaction of ions with their URWs in kicker undulators.

Below we accept a Gaussian distribution for the URW, its Rayleigh length
$Z_R = 4\pi \sigma _w^2 / \lambda _{1\min } = L_u / 2$, the rms waist size
$\sigma _w^ = \sqrt {L_u \lambda _{1\min } / 8\pi } $. In this case the rms
electric field strength $E_w $ of the wavelet in the kicker undulator

\begin{equation}
\label{eq4}
E_w = \sqrt {2\Delta E_1 / \sigma _w^2 \lambda _{1\min } } = 8\sqrt
{\pi} r_i \gamma ^3\sqrt {\overline {B^2} } / L_u (1 + K^2)^{3 /
2}.  \end{equation}

The rate of the energy loss for ions in the amplified URW is

\[
P_{loss} = eZE_w L_u \beta _{ \bot m} f N_{kick} \sqrt {\alpha _{ampl} } =
\]
\begin{equation}
\label{eq5}
8\sqrt {\pi} eZr_{i} f {\gamma ^{2}} N_{kick} \sqrt {\alpha _{ampl}
\cdot \overline {B^2} } K / (1 + K^2)^{3 / 2}, \end{equation}

\noindent
where $\beta _ \bot = K / \gamma $; $f$ is the revolution frequency;
$N_{kick} $ is the number of kicker undulators; $\alpha _{ampl} $ is the
gain in optical amplifier.

The damping time for the ion beam in the longitudinal degree of freedom is

\begin{equation}
\label{eq6}
\tau = \sigma _E / P_{loss} ,
\end{equation}

\noindent
where $\sigma _E $is the energy spread of the ion beam.

According to (\ref{eq6}), the damping time for EOC is proportional to the energy
spread of the beam which is much less then the energy of ions included in
similar expression for damping time controlled by Robinson's damping
criterion. Moreover, because of the non-exponential decay of both energy and
angular spreads of the beam the degree of cooling of ion beams for EOC is
much higher than 1/e reduction of these parameters.

Note that the higher the dispersion function and the less the beta function
at the location of the kicker undulator the higher the rate of damping of
betatron oscillations. In this case low energy jumps of ions lead to large
jumps of closed orbits and near the same large jumps of betatron oscillation
amplitudes.

\section{STOCHASTIC PROCESSES IN THE EOC }

URW of one ion does not disturb trajectories of other ions if an average
distance between ions in the longitudinal direction is more, than the URW's
length, $M\lambda _{UR,1} $, and the transverse dimensions of the URW's in
kicker undulators are overlapped and higher then the transverse total
(dispersion + betatron) dimensions of the being cooled ion beam. This case
is named ``single ion in the sample''. It corresponds to the beam current

\begin{equation}
\label{eq7}
i < i_c = \frac{Zec}{M\lambda _{1\min } } = \frac{4.8 \cdot 10^{ -
9}Z}{M\lambda _{1\min } [cm]}[A].
\end{equation}

If $i > i_c $ amplified URWs do not disturb the energy spread and amplitudes
of betatron oscillations of other's ions of the beam in the first
approximation and change them in the second one because of a stochasticity
of the initial phases of ions in other's URWs. Stochasticity limits the
degree of cooling.

Open ions of the beam loose their energies up to the moment when they will
be displayed inward to the distances corresponding to overlapping their URWs
by the stopped screen. After this time all ions will stay at the threshold
energy with the equilibrium energy spread and the spread of amplitudes of
betatron oscillations [3]

\begin{equation}
\label{eq8}
\left( {\sqrt {\overline {A_x^2 } } } \right)_{eq} = \left( {\sqrt
{\overline {x_\eta ^2 } } } \right)_{eq} = \frac{1}{2}\vert \delta x_\eta
\vert (n_c + 1 + n_n )
\end{equation}

\noindent
determined by the average jump of the ion energy $\Delta E = P_{loss} \sqrt
{n_c + 1 + n_n } / f$, where $\delta x_\eta = \eta _x \beta ^{ -
2}(P_{loss} / Ef)$, $n_c = i / i_c $ is the number of ions in a sample;
$n_n = N_n / N_{ph} $; $N_n $, the number of noise photons in the URW sample
at the amplifier front end.

\section{STORAGE RING LATTICE REQUIREMENTS }

The relative phase shifts of ions in their URWs radiated in the pickup
undulator and displaced to the entrance of kick undulators depend on theirs
energy and amplitude of betatron oscillations. If we assume that the
longitudinal shifts of URWs $\Delta l < \lambda _{UR} / 2$, then the
amplitudes of betatron oscillations, transverse horizontal emittance of the
beam, in the smooth approximation, and the energy spread of the beam must
not exceed the values

\[
A_x < < A_{x,\lim } = \frac{\lambda _{1\min } \sqrt {\lambda _{bet} } }{\pi
},
\quad
\varepsilon _x < 2\lambda _{UR1\min } ,
\]
\begin{equation}
\label{eq9}
\frac{\Delta \gamma }{\gamma } < (\frac{\Delta \gamma }{\gamma })_{\lim } =
\frac{\beta ^2}{\eta _c }\frac{\lambda _{1\min } }{\lambda _{bet} },
\end{equation}

\noindent
where $\lambda _{x,bet} = C / v_x $; C is the circumference of the ring,
$v_x $, the tune; $\eta _c = \alpha _c - \gamma ^{ - 2}$ and $\alpha _c $
are local slip and momentum compaction factors between undulators.

Strong limitations (\ref{eq9}) to the energy spread can be overcame if, according to
the decrease of the high energy edge of the being cooled beam, a change in
time of optical paths of URWs is produced. Special elements in storage ring
lattices (short inverted dipoles, quadrupole lenses et al.) to decrease the
slip [5-8] can be used as well. With cooling of fraction of the beam at a
time only, the lengthening problem diminishes also as the $\Delta E / E$ now
stands for the energy spread in the part of the beam which is under cooling
at the moment.

\section{OPTICAL SYSTEM FOR EOC}

The power of the optical amplifier is equal to the power of the amplified
URWs plus the noise power

\begin{equation}
\label{eq10}
P_{ampl} = \varepsilon _{sample} \cdot f \cdot N_i + P_n ,
\end{equation}

\noindent
where $\varepsilon _{sample} = \hbar \omega _{1,\max } N_{ph} \alpha _{ampl}
$ is the energy in a sample; $N_i $, the number of ions in the ring. The
bunch spacing in LHC (45 m) is much bigger than the bunch length ($\sim
$5-10 cm). The same time structure of the OPA must be used. The energy of
OPA emitted for the damping time is proportional to the initial energy
spread of the ion beam. It can be decreased by decreasing peak RF voltage or
by increasing the number of bunches.

The space resolution of the ion beam is [3]

\begin{equation}
\label{eq11}
\delta x_{res} \simeq 1.22\lambda _{1\min } / \Delta \theta = 1.22\sqrt
{\lambda _{1\min } L_u }.
\end{equation}

The transverse selectivity of radiation (movable screen) can be arranged
with help of electro-optical elements. These elements contain crystals,
which change their refraction index while external voltage applied. This
technique is well known in optics [9]. In simplest case the sequence of
electro-optical deflector and a diaphragm followed by optical lenses, allow
controllable selection of radiation generated by different parts of the
beam.

\textit{Example 1.}\textsf{ }EOC of fully stripped $_{207}^{82} Pb$ ion beam in the
CERN LHC at the injection energy $M_i c^2\gamma =  \quad  = 36.9$TeV. The
parameters of the LHC: circumference C=27 km, $f = 1.1 \cdot 10^4$, $v_x =
64.28$, $\alpha _c = 3.23 \cdot 10^{ - 4}$, $i = 6.28$mA, $\eta _c = 3.18
\cdot 10^{ - 4}$,$^{ }N_i = 4.1 \cdot 10^{10}$, $\gamma = 190.5$, $\Delta
\gamma / \gamma = 3.9 \cdot 10^{ - 4}$, normalized emittance $ \in _{x,n} =
1.4_ \mu m$, beta and dispersion functions at the kicker undulator $\beta _x
= 25.0$ m, $\eta _x = 2.0$ m, betatron beam size at pickup undulator $\sigma
_{x,0} = 0.43$mm, dispersion beam size $\sigma _{\eta ,0} = 0.95$ mm, total
beam size $\sigma _{b,0} = 1.1$ mm.

One pickup one kick helical undulator with parameters $\sqrt {\overline
{B^2} } = 10^5$Gs, $\lambda _u = 4$ cm, $M = $300 and two optical parametric
amplifiers (OPA) with gains $\sim $10$^4$ are used. The total gain goes to
be $\alpha _{ampl} = 10^8$.

In this case: $M_i c^2 = 1.94 \cdot 10^{11}$ eV, $N_{ph} = 1.01 \cdot 10^{ -
2}$, $r_i = 4.96 \cdot 10^{ - 15}$cm, $i_c = 0.024$mA, $N_c = i_c / ef =
1.7 \cdot 10^8$, $\lambda _{1\min } = 5.5 \cdot 10^{ - 5}$ cm, $\sigma _w
= 0.51$ mm, $K = 0.0081$,$E_w \cong  \quad 1.22 \cdot 10^{ - 2}$ V/cm, $P_{loss}
= 5.61 \cdot 10^6$ eV/sec, $\tau = 42.8$ min, $P_{ampl} = 164$ W, $\delta
x_\eta = 2.76 \cdot 10^{ - 9}$ cm, $\lambda _{x,bet} = 414.7$ m, $A_{x,\lim
} = 5$mm, $(\Delta \gamma / \gamma )_{\lim } = 4.17 \cdot 10^{ - 6}$,
$\delta x_{res} = 3.14$mm, $n_c = 241$, $n_n = 99$, equilibrium beam
dimensions $\left( {\overline {A_x^2 } } \right)_{eq}^{1 / 2} =  \quad \left(
{\overline {x_\eta ^2 } } \right)_{eq}^{1 / 2} = 3.45 \cdot 10^{ - 7}$ cm,

Damping time of an ion beam, according to (\ref{eq5}), (\ref{eq6}), is $\tau = \sigma _E /
P_{loss} \vert _{K < 0.5}^ \sim \sigma _E / N_{kick} K\gamma ^2$. If
$\lambda _{1\min } $, $\sigma _E $ and $\overline {B^2} $ are constants,
then $\tau \sim 1 / \gamma ^4$ and the power of the optical amplifier
$P_{ampl} \sim N_{ph} N_i \sim N_i \gamma ^4$ Damping time can be decreased
by using beams with smaller initial energy spread $\sigma _E $, many kicker
undulators $N_{kick} > 1$ and higher gain of optical amplifiers.

\textit{Example 2.} EOC of fully stripped $_{207}^{82} Pb$ ion beam in the CERN LHC at the
energy $\gamma = 953$, $\Delta \gamma / \gamma =  \quad 1.3 \cdot 10^{ - 4}$ ,
$N_i = 10^8$, betatron beam size $\sigma _{x,0} = 4.3$mm, dispersion beam
size $\sigma _{\eta ,0} = 9.5$ mm, total beam size $\sigma _{b,0} = 11$ mm
at pickup undulator. One pickup and one kick undulator with parameters
$\sqrt {\overline {B^2} }  \quad  = 10^5$Gs, $\lambda _u = 1$ m, $K = 0.202$, $M =$12
and OPAs identical to ones of the example 1 are used.

In this case: $N_{ph} = 6.29$, $\lambda _{1\min } = 5.5 \cdot 10^{ - 5}$ cm,
$\sigma _w =  \quad 0.51$ mm, $\delta x_{res} = 3.14$ mm, $E_w \cong 1.44$ V/cm,
$P_{loss} =  \quad 3.3 \cdot 10^9$ eV/sec, $P_{ampl} = 249$ W, $\delta x_\eta =
3.25 \cdot 10^{ - 6}$ cm, $\tau = 4.36$ sec, the equilibrium beam dimensions
$\left( {\overline {A_x^2 } } \right)_{eq}^{1 / 2} =  \quad \left( {\overline
{x_\eta ^2 } } \right)_{eq}^{1 / 2} = 1.22 \cdot 10^{ - 6}$ cm.

In these examples initial transverse beam dimensions $\sigma _{x,0} <
A_{x,\lim } $. It means that there is no problem with the dependence of
phase shifts of ions on their amplitudes. At the same time there is a
necessity in special elements in storage ring lattices to decrease the slip.

We were forced to decrease the number of ions in the second example to reach
acceptable power of OPA. It can be increased if the gain of OPA will be
decreased. At that damping time will be increased.

The space resolution of the ion beam and that is why its final dimensions
are limited by a value $\delta x_{res} = 3.14$ mm, which is larger then the
initial beam dimension $\sigma _{b,0} = 1.1$ mm in the example 1. It does
not permit to reach small equilibrium beam dimensions and means that we must
increase the initial beam dimensions by increasing the dispersion and beta
functions in the location of the pickup undulator or, according to (11),
decrease the length of the undulator and using more short $\lambda _{1\min }
$ to increase the resolution. High beam size in the pickup undulator can be
arranged by appropriate betatron function at this place.

Cooling of ion beams is produced in the RF bucket. In this case the screen
must be moved to the position of the image of the equilibrium orbit and
stopped at this position. Cooling cycles must be repeated periodically. The
length of the URW bunch can be less then ion one. In this case only
overlapped parts of ion and URW beams will interact. Despite of this all ion
beam will be cooled as ions take part in phase oscillations.

\section{CONCLUSION}

We considered EOC of ion beams in LHC storage ring. Details of EOC for
bunched beam will be presented in a separate paper. The gain of the OPAs can
be higher then considered in the paper. That is why the damping time is
limited by the maximum achievable power of the OPA and number of kicker
undulators.


\begin{thebibliography}{xx}

\bibitem{1}
A.Mikhailichenko, M.Zolotorev, Phys.Rev.Lett.71:
\textsf{4146-4149,1993.}

\bibitem{2}
E.G.Bessonov, physics/0404142.

\bibitem{3}
E.G.Bessonov, A.A.Mikhailichenko, Proc. PAC05, May 16-20, 2005,
Knoxville, Tennessee, USA. 
http://accelconf.web.cern.ch/accelconf/p05/PAPERS/
TPAT086.PDF.

\bibitem{4}
E.G.Bessonov, A.A.Mikhailichenko, A.V.Poseryaev, Physics of the Enhanced
optical cooling of particle beams in storage rings, 
http://arxiv.org/abs/physics/0509196.

\bibitem{5}
C.Pellegrini and D.Robin, Nucl. Instr. Meth. A301,
27 (1991); Proc. IEEE Part. Accel. Conf., San
Francisco, p. 398 (1991).

\bibitem{6}
M.Berz, Part. Acc. V.24, p.109, 1989.

\bibitem{7}
E.Forest, M.Berz, J.Irwin, Part. Acc. V.24, p.91 (1989).

\bibitem{8}
Klaus G.Steffen, High Energy Beam Optics,
Interscience publishers, NY-London-Sydney, 1965.

\bibitem{9}
V.J.Fowler, J.Schlafer, Applied Optics Vol.5, N10, 1657 (1966).

\end{thebibliography}
\end{document}